\documentclass[12pt,journal,onecolumn]{IEEEtran}
\usepackage{epsfig,rotating,setspace,latexsym,amsmath,epsf,amssymb,amsfonts,bm,theorem,subfigure,epstopdf}
\usepackage{cite,authblk}
\usepackage{algorithmic,algorithm}
\usepackage{setspace}
\usepackage{color}
\newtheorem{definition}{Definition}

\IEEEoverridecommandlockouts

\allowdisplaybreaks

\begin{document}

\title{Energy Harvesting Communications under Explicit and Implicit Temperature Constraints\thanks{This work was supported by NSF Grants CNS 13-14733, CCF 14-22111, and CNS 15-26608; and will be presented in part at IEEE Global Communications Conference, Singapore, December 2017.}}

\author[1]{Abdulrahman Baknina}
\author[2]{Omur Ozel}
\author[1]{Sennur Ulukus}
\affil[1]{\normalsize Department of Electrical and Computer Engineering, University of Maryland, MD}
\affil[2]{\normalsize Department of Electrical and Computer Engineering, Carnegie Mellon University, PA}

\maketitle

\vspace*{-0.4cm}

\begin{abstract}
We consider an energy harvesting communication system where the temperature dynamics is governed by the transmission power policy. Different from the previous work, we consider a discrete time system where transmission power is kept constant in each slot. We consider two models that capture different effects of temperature. In the first model, the temperature is constrained to be below a critical temperature at all time instants; we coin this model as {\it explicit temperature constrained model}. We investigate throughput optimal power allocation for multiple energy arrivals under general, as well as temperature and energy limited regimes. We show that the optimal power allocation for the temperature limited case is monotone decreasing. In the second model, we consider the effect of the temperature on the channel quality via its influence on additive noise power; we coin this model as {\it implicit temperature constrained model}. In this model, the change in the variance of the additive noise due to previous transmissions is non-negligible. In particular, transmitted signals contribute as interference for all subsequent slots and thus affect the signal to interference plus noise ratio (SINR). In this case, we investigate throughput optimal power allocation under general, as well as low and high SINR regimes. We show in the low SINR regime that the optimal allocation dictates the transmitter to save its harvested energy till the last slot. In the high SINR regime, we show that the optimal power sequence is monotone increasing. Finally, we consider the case in which implicit and explicit temperature constraints are simultaneously active and we show under certain conditions that the optimal power sequence is monotone decreasing.
\end{abstract}

\section{Introduction}

We consider two different effects of temperature on the optimal power allocation in a single-user energy harvesting communication system. These effects show themselves as explicit and implicit temperature constraints on the power allocation. In the explicit temperature constrained model, a peak temperature constraint prevents the system from overheating. In the implicit temperature constrained model, the effect of the temperature on the channel quality controls the system temperature.

The scheduling-theoretic approach for energy harvesting communications was studied in various settings, see \cite{jingP2P, kayaEmax, omurFade, ruiZhangEH, jingBC, jingMAC, aggarwalPmax, kaya-interference,orhan2015energy, gunduz2hop, huang2013throughput, varan_twc_jour, arafa2017energy, gurakan2016cooperative, gunduzLoss, orhan-broadband, ruiZhangNonIdeal, omurHybrid, kayaLoss, kayaRxEH, yatesRxEH1, yatesRxEH2, yatesRxEH3, payaroRxEH, arafaJSACdec, omur_temp_journ, bakninaenergy}. Previous works considered single-user channel \cite{jingP2P, kayaEmax, omurFade, ruiZhangEH}, broadcast channel \cite{jingBC}, multiple access channel \cite{jingMAC, aggarwalPmax}, interference channel \cite{kaya-interference}, two-hop channel \cite{orhan2015energy,gunduz2hop, huang2013throughput}, two-way channel \cite{varan_twc_jour, arafa2017energy}, and diamond channel \cite{gurakan2016cooperative}. The effect of imperfect transmitter circuitry is considered in \cite{gunduzLoss, orhan-broadband, ruiZhangNonIdeal, omurHybrid, kayaLoss}. Receiver side energy harvesting communication systems is considered in \cite{kayaRxEH, yatesRxEH1, yatesRxEH2, yatesRxEH3, payaroRxEH, arafaJSACdec}. Temperature effects are studied in \cite{omur_temp_journ, bakninaenergy}. In \cite{omur_temp_journ}, a peak temperature constraint is considered and the optimal continuous power allocation is studied. Although extensive insights and properties of the optimal policy were derived, only the single energy arrival case was fully solved.
In this paper, we consider a discrete time version of the problem considered in \cite{omur_temp_journ, bakninaenergy}.

In the first model we consider here, which we coin as the {\it explicit temperature constrained model}, we consider an explicit peak temperature constraint. Increasing the transmission power increases the throughput and the temperature. Higher temperature levels mean smaller admissible transmission power levels for future slots. This model is the slotted version of \cite{omur_temp_journ}. We consider the optimal power allocation for the multiple energy arrival case and we determine a generalized water-filling solution. When the temperature constraint is not binding, the problem reduces to the single-user energy harvesting channel studied in \cite{jingP2P}, where the optimal power sequence is monotone increasing. When the energy constraint is not binding, we show that the optimal power sequence is monotone decreasing, and the resulting temperature is monotone increasing.

In the second model we consider here, which we coin as the {\it implicit temperature constrained model}, the temperature is not explicitly constrained, however, the temperature affects the additive noise power and hence the channel quality. This arises when the dynamic range of the temperature is large, and is similar to that presented in \cite{koch2009channels} but in a scheduling-theoretic setting. In this case, the transmit powers used in previous channel uses affect the thermal noise and therefore the channel quality in future channel uses, and hence, the channel becomes a {\it use dependent} or {\it action dependent} channel, see \cite{ward2015optimal, weissman2010capacity, ahmadi2012channels}.

In the implicit temperature constrained model, transmissions in the previous slots interfere with the current transmission due to temperature dependent noise and the causality of the temperature filter. This filter is the discrete time version of the continuous time first order filter that defines the temperature dynamics. For the general signal to interference plus noise ratio (SINR), we observe that the problem is non-convex and is a signomial problem for which we obtain a local optimal solution using the single condensation method in \cite{chiang2007power}. We then propose a heuristic algorithm which improves upon the local optimal solution and may achieve the global optimal solution. Then, we consider the extreme settings of low and high SINR regimes. We show that in the low SINR regime, saving energy till the last slot and transmitting only in the last slot is optimal. For the high SINR regime, we observe that the problem is a geometric program and we explore specific structural results in this setting. Expanding upon the equivalence of this problem to its convex counterpart via a one-to-one transformation, we show that the KKT conditions in the original problem have a unique solution. Then, we obtain an algorithm to solve the KKT conditions in the original problem. We show convergence of this algorithm to the unique solution of the KKT conditions. We then show that for this unique solution, the power sequence is monotone increasing; hence, proving the monotone increasing property of the optimal power sequence.

Finally, we consider the case when implicit and explicit temperature constraints are simultaneously active. In general, we observe that the problem is non-convex and the same signomial programming approach as in the implicit temperature constrained case is applicable. For high SINR case, the problem is a geometric program and we show in the temperature limited case that the optimal power sequence is monotone decreasing under certain conditions.

\section{System Model}

\begin{figure}[t]
	\centerline{\includegraphics[width=0.5\columnwidth]{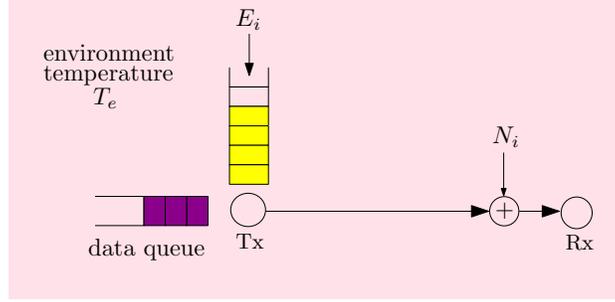}}
	\caption{System model: the system heats up due to data transmission.}
	\label{sys_model}
\end{figure}

We consider an energy harvesting communication system in which the transmitter harvests energy $\tilde{E}_i$ in the $i$th slot, see Fig. \ref{sys_model}.
We consider the temperature model considered in \cite{omur_temp_journ,bakninaenergy}. In this model, the temperature, $T(t)$, evolves according to the following differential equation,
\begin{align}
\label{eq_penne_heat}
\frac{dT(t)}{dt}=a p(t) - b (T(t)-T_e)
\end{align}
where $T_e$ is the environment temperature, $T(t)$ is the temperature at time $t$, $p(t)$ is the power, and $a,b$ are non-negative constants. With the initial temperature $T(0)=T_e$, the solution of (\ref{eq_penne_heat}) is:
\begin{align} \label{temp-exp}
T(t)= e^{-bt} \int_{0}^{t} e^{b \tau} a p(\tau )  d \tau +T_e
\end{align}

In what follows we assume that the duration of each slot is equal to $\Delta$, which can take any positive value.
Let us define $T_i \triangleq T(i \Delta)$ as the temperature level by the end of the $i$th slot, $P_i \triangleq P(i \Delta)$ as the power level used in the $i$th slot. Using (\ref{temp-exp}), $T_i$ can be expressed as:
\begin{align}
T_{i}  =& e^{-bi\Delta} \int_{0}^{i\Delta} e^{b \tau} a p(\tau )  d \tau  +T_e \\
=& e^{-b\Delta}   e^{-b(i-1)\Delta} \int_{0}^{(i-1)\Delta} e^{b \tau} a p(\tau )  d \tau   +
e^{-bi\Delta} \int_{(i-1)\Delta}^{i\Delta} e^{b \tau} a P_{i}  d \tau  +T_e  \label{eq_temp_monotone}\\
=& e^{-b\Delta}   (T_{i-1}-T_e)  +
\frac{  a P_{i}}{b}  \left[ 1 -e^{-b\Delta} \right]   +T_e \\
=& \alpha T_{i-1} + \beta P_i +\gamma \label{eq_temp_discrete}
\end{align}
where $\alpha =e^{-b\Delta}$, $\beta = \frac{a}{b}\left[1- \alpha \right]$ and $\gamma =T_e \left[1- \alpha \right]$.

The effect of $\Delta$ in (\ref{eq_temp_discrete}) appears through the constants $\alpha,\beta, \gamma$. As the slot duration increases, the values of $\beta, \gamma$ increase while the value of $\alpha$ decreases; as the slot duration increases, the temperature at the end of the slot becomes more dependent on the power transmitted within this slot and less dependent on the initial temperature at the beginning of the slot.

We now eliminate the previous temperature readings in $T_i$ making the temperature a function of the powers only. We can do this by recursively substituting $T_{i-1}$ in $T_i$ in (\ref{eq_temp_discrete}) to have
\begin{align}
T_k=  \beta \sum_{i=1}^{k} \alpha^{k-i} P_{i} + T_e \label{eq_temp_in_terms_of_power}
\end{align}
This formula shows that the temperature at the end of each slot depends on the power transmitted in this slot and all previous slots through an exponentially decaying \emph{temperature filter}. We note that this is the same formula that was developed in \cite{koch2009channels} in which the slot duration was assumed to be unity; here we assume a general slot duration which is equal to $\Delta$. In what follows, we denote the vector of elements by the bold letter without a subscript, i.e., for example, the vector of powers is defined as $\mathbf{P}\triangleq[P_1, \ldots, P_D]$.

\section{Explicit Peak Temperature Constraint}\label{sect_explicit}

We now consider the model in which we have an energy harvesting transmitter with a peak temperature constraint. The noise variance is the same throughout the communication session and is set to $\sigma^2$. We consider a slotted system with a constant power per slot. There are $D$ slots. It follows from (\ref{eq_temp_monotone}) (and also \cite[equation (47)]{omur_temp_journ}), that the temperature is monotone within the slot duration. Hence, for the peak temperature constrained case, it suffices to constrain the temperature only at the end of each slot; we begin the communication with the system having temperature $T_e$. In this case, the problem can be written as
\begin{align}
\max_{\mathbf{P}\geq \mathbf{0}} \quad & \sum_{i=1}^{D} \frac{\Delta}{2}  \log \left(1+\frac{P_i}{\sigma^2} \right)     \nonumber \\
\mbox{s.t.} \quad &  T_{k} \leq T_c \nonumber\\ & \sum_{i=1}^{k}  \Delta P_{i} \leq \sum_{i=1}^{k} \tilde{E}_i, \ \forall k  \label{temp_and_energy_intermsof_energy}
\end{align}
where $\Delta$ in the objective function and the energy constraint is to account for the slot duration. In what follows, without loss of generality, we drop $\Delta$ since it is just a constant multiplied in the objective function and by defining $E_i= \frac{\tilde{E}_i}{\Delta} $.

We rewrite problem (\ref{temp_and_energy_intermsof_energy}) making use of (\ref{eq_temp_in_terms_of_power}) as
\begin{align}
\max_{\mathbf{P}\geq \mathbf{0} } \quad & \sum_{i=1}^{D} \frac{1}{2}  \log \left(1+\frac{P_i}{\sigma^2} \right)     \nonumber \\
\mbox{s.t.} \quad &  \sum_{i=1}^{k} \alpha^{k-i} P_{i} \leq \frac{T_c-T_e }{\beta} \nonumber\\ &  \sum_{i=1}^{k}  P_{i} \leq \sum_{i=1}^{k} E_i, \ \forall k \label{temp_and_energy}
\end{align}
In the last slot, either the temperature or the energy constraint has to be satisfied with equality. Otherwise, we can increase one of the powers until one of the constraints is met with equality and this strictly increases the objective function.

This problem is a convex problem in the powers, which can be solved optimally using the KKTs. The Lagrangian is:
\begin{align}
\mathcal{L}= & - \sum_{i=1}^{D}    \log \left(1+\frac{P_i}{\sigma^2} \right)    + \sum_{k=1}^{D} \lambda_k \left( \sum_{i=1}^{k} \alpha^{k-i} P_{i} - \frac{T_c-T_e }{\beta}\right)  + \sum_{k=1}^{D} \mu_k \left(\sum_{i=1}^{k}  P_{i} - \sum_{i=1}^{k}  E_{i}  \right)
\end{align}
Differentiating with respect to $P_i$ and equating to zero we get,
\begin{align}
P_i =\frac{1}{\alpha^{-i} \sum_{k=i}^{D} \lambda_k \alpha^k  +\sum_{k=i}^{D}\mu_k} -\sigma^2 \label{eq_optimal_powers_both_E_T}
\end{align}
In the optimal solution, if neither constraint was tight in slot $i<D$, then the power in slot $i+1$ is strictly less than the power in slot $i$.
This follows from complementary slackness since if at slot $i$, if both constraints were not tight then we have $\lambda_i=\mu_i=0$ which, using (\ref{eq_optimal_powers_both_E_T}), implies that $P_{i}>P_{i+1}$.

To find the optimal solution, we begin with an initial feasible power allocation. If for this power allocation there exist non-negative Lagrange multipliers which satisfy (\ref{eq_optimal_powers_both_E_T}) and the complementary slackness conditions, then this is the optimal power allocation. Otherwise, these power allocations should be modified by pouring water away from the slots with negative Lagrange multipliers to the slots with higher Lagrange multipliers until non-negative Lagrange multipliers are found and the complementary slackness conditions are satisfied. Since this problem is a convex optimization problem, any solution for the KKTs achieve the global maximum.

\subsection{Energy Limited Case}
In this subsection, we study a sufficient condition under which the system becomes energy limited. For all slots $j$ in which the following is satisfied
	\begin{align}\label{eq_tner_lim_cond}
	\sum_{i=1}^{j} E_i \leq \frac{T_c-T_e }{\beta}
	\end{align}
	the temperature constraint cannot be tight. Intuitively, in this case, the incoming energy is so small that it can never overheat the system. Therefore, the binding constraint here is the availability of energy. In particular, when (\ref{eq_tner_lim_cond}) is satisfied for $j=D$, then the temperature constraint can be completely removed from the system.
To prove this, we assume for the sake of contradiction that we have at slot $j$, $\sum_{i=1}^{j} E_i \leq \frac{T_c-T_e }{\beta}$ while the temperature constraint is tight, which implies:
	\begin{align}
	\frac{T_c-T_e }{\beta}= \sum_{i=1}^{j} \alpha^{j-i} P_{i} < \sum_{i=1}^{j}  P_{i}   \leq \sum_{i=1}^{j} E_i
	\end{align}
	which contradicts the assumption $\sum_{i=1}^{j} E_i \leq \frac{T_c-T_e}{\beta}$. The strict inequality follows since $\alpha<1$. The structure of the optimal solution for this case is studied in \cite{jingP2P}.

\subsection{Temperature Limited Case}\label{sec_temp_limited}
In this subsection, we first study a sufficient condition for problem (\ref{temp_and_energy}) to be temperature limited.
The energy constraint is never tight if the following condition is satisfied:
	\begin{align}\label{eq_cond_temp_limited}
	\frac{T_c-T_e}{\beta} < \frac{\sum_{i=1}^{k}   E_i}{k}, \ \forall k \in \{1, \ldots, D\}
	\end{align}
	Intuitively, the incoming energy is so large that there will never be a shortage of energy. Therefore, the binding constraint here is overheating the system.
For the temperature limited case, an upper bound on the transmission powers is equal to $\frac{T_c-T_e}{\beta}$. This follows because for any slot $k$ we have $\sum_{i=1}^{k-1} \alpha^{k-i} P_{i}+P_k \leq \frac{T_c-T_e }{\beta}$, thus $P_k$ can be at most equal to $\frac{T_c-T_e}{\beta}$. Hence, (\ref{eq_cond_temp_limited}) is sufficient to satisfy $\sum_{i=1}^{k}   P_i < \sum_{i=1}^{k} E_i$.

In what follows, we study the structure of the optimal policy for the temperature limited case.
In the last slot, the temperature constraint is satisfied with equality. The optimal powers are monotonically decreasing in time.
	The proof follows by contradiction. Assume for some index $j$ that we have $P_j^* <  P_{j+1}^*$. We now form another policy, denoted as $\{\bar{P}_i\}$, which has $\bar{P}_i = P_i^*$ for all slots $i \neq j, j+1 $, while we change the powers of slots $j,j+1$ to be $\bar{P}_j = P_{j}^*+\delta$ and $
	\bar{P}_{j+1} = P_{j+1}^*-\delta$ for small enough $\delta>0$. This $\delta$ always exists as $P_j^* <  P_{j+1}^*$ implies that  $\sum_{k=1}^{j} \alpha^{j-k} P_k^* < \frac{T_c-T_e}{\beta}$. Since the objective function is strictly concave, this new policy yields a strictly higher objective function, which contradicts the optimality of $P_j^* <  P_{j+1}^*$.
	Now it remains to check that with this new policy, the temperature constraint is still feasible for any slot $k \geq j+1$ which follows from:
	\begin{align}
	\sum_{i=1, \neq j,j+1}^{k} \alpha^{k-i} \bar{P}_{i}  + \alpha^{k-j} \bar{P}_{j} +\alpha^{k-j-1} \bar{P}_{j+1}
	&= \sum_{i=1, \neq j,j+1}^{k} \alpha^{k-i} P_{i}^*
	+ \alpha^{k-j} \bar{P}_{j} +\alpha^{k-j-1} \bar{P}_{j+1} \\
	&<  \sum_{i=1, \neq j,j+1}^{k} \alpha^{k-i} P_{i}^*
	+ \alpha^{k-j} {P}_{j}^* +\alpha^{k-j-1} {P}_{j+1} ^* \\
	& = \sum_{i=1}^{k} \alpha^{k-i} P_{i}^* \\ &< \frac{T_c-T_e }{\beta}
	\end{align}
	Since this is valid for any $k \geq j+1$, we can take in particular $k=D$. Now we can increase any of the powers to satisfy the last inequality by equality which strictly improves the objective function. Hence, this violates the optimality of any policy which has $P_i^* <  P_{i+1}^*$ for any $i\in \{1,\ldots,D\}$.

 	Moreover, the optimal temperature levels are non-decreasing in time.
 	To prove this, using (\ref{eq_temp_in_terms_of_power}), it suffices to show that:
 	\begin{align}
 	\sum_{i=1}^{k} \alpha^{k-i} P_{i}^* \leq \sum_{i=1}^{k+1} \alpha^{k+1-i} P_{i}^*, \ \ \forall k=\{1,\ldots, D-1\} \label{temp_inc_eq_1}
 	\end{align}
 	We rewrite (\ref{temp_inc_eq_1}) as follows,
 	\begin{align}
 	(1-\alpha)\sum_{i=1}^{k} \alpha^{k-i} P_{i}^* \leq   P_{k+1}^*, \ \ \forall k=\{1,\ldots, D-1\} \label{temp_inc_eq_2}
 	\end{align}
 	Since, we know that the last slot has to be satisfied with equality then we know $\sum_{i=1}^{D} \alpha^{D-i} P_{i}^*=\frac{T_c-T_e }{\beta}$. Hence, for the constraint at $k=D-1$ we have:
 	\begin{align}
 	\sum_{i=1}^{D-1} \alpha^{D-1-i} P_{i}^* \leq \frac{T_c-T_e }{\beta}=\sum_{i=1}^{D} \alpha^{D-i} P_{i}^* \label{temp_inc_eq_3}
 	\end{align}
 	which can be written as follows
 	\begin{align}
 	(1-\alpha)\sum_{i=1}^{D-1} \alpha^{D-1-i} P_{i}^* \leq   P_{D}^* \label{temp_inc_eq_4}
 	\end{align}
 	which proves (\ref{temp_inc_eq_2}) for $k=D-1$. Now assume for the sake of contradiction that (\ref{temp_inc_eq_2}) is false for $k=D-2$, i.e.:
 	\begin{align}
 	P_{D-1}^*< (1-\alpha ) \sum_{i=1}^{D-2} \alpha^{D-2-i} P_{i}^*
 	\end{align}
 	Substituting this in (\ref{temp_inc_eq_4}), we get:
 	\begin{align}
 	P_{D-1}^* &=  \alpha P_{D-1}^*  + (1-\alpha)P_{D-1}^* \\
 	&<\alpha (1-\alpha)\sum_{i=1}^{D-2} \alpha^{D-2-i} P_{i}^* +(1-\alpha)P_{D-1}^* \\
 	&=(1-\alpha)\sum_{i=1}^{D-1} \alpha^{D-1-i} P_{i}^*  \leq   P_{D}^*
 	\label{temp_inc_eq_5}
 	\end{align}
 	But since we know that in the optimal policy the power sequence is monotone decreasing, this is a contradiction and (\ref{temp_inc_eq_2}) holds for $k=D-2$.
 	The same argument follows for any $k<D-2$.

	In the optimal solution, if the constraint is satisfied with equality for two consecutive slots then the power in the second slot must be equal to $(1-\alpha) \frac{T_c-T_e}{\beta}$.
	To obtain this, the two consecutive constraints which are satisfied with equality are solved simultaneously for the power in the second slot. In addition, when the temperature hits the critical temperature for the first time, the transmission power in that slots will be \emph{strictly higher} than $(1-\alpha) \frac{T_c-T_e}{\beta}$.
	To show this we denote the time slot at which the temperature hits $T_c$ for the first time as $i^*$. Hence, we have:
	\begin{align}
	\sum_{i=1}^{i^*-1} \alpha^{i^*-1-i} P_i < \frac{T_c-T_e}{\beta} , \ \ \sum_{i=1}^{i^*} \alpha^{i^*-i} P_i = \frac{T_c-T_e}{\beta}
	\label{eq_strict_temp_1}
	\end{align}
	Using both equations in (\ref{eq_strict_temp_1}) simultaneously we have:
	\begin{align}
	(1-\alpha) \frac{T_c-T_e}{\beta} < P_{i^*}
	\end{align}
	which is the power of the slot at which temperature hits the critical temperature for the first time.
	
Hence, when the temperature hits the critical temperature, the optimal transmission power in all the subsequent slots becomes constant and equal to $(1-\alpha) \frac{T_c-T_e}{\beta}$.
This follows since the temperature is increasing, thus whenever the constraint becomes tight, it remains tight for all subsequent slots. We now conclude that the transmission power at all slots are bounded as follows
\begin{align}
(1-\alpha) \frac{T_c-T_e}{\beta} \leq P_i \leq   \frac{T_c-T_e}{\beta} , \ \ \forall i=\{1,\ldots, D\}
\end{align}
The lower bound follows from the discussion above while the upper bound follows from the feasibility of the constraints.

We now proceed to find the optimal power allocation. Since the problem is convex, a necessary and sufficient condition is to find a solution satisfying the KKTs. The optimal power is given by setting $\boldsymbol{\mu}=\mathbf{0}$ in (\ref{eq_optimal_powers_both_E_T}), which gives:
\begin{align}
P_i =\frac{\alpha^{i}}{ \sum_{k=i}^{D} {\lambda}_k \alpha^k } - \sigma^2 \label{eq_optimal_powers_T}
\end{align}
It follows from the complementary slackness that if at slot $i$ the temperature constraint is satisfied with strict inequality then $P_{i+1}<P_i$.

To this end we present an approach to obtain the optimal powers. We use line search to search for the time slot at which the temperature constraint becomes tight, which we denote as $i^*$. Then, slots $i=\{i^*+1, \ldots, D\}$ have power allocation equal to $(1-\alpha) \frac{T_c-T_e}{\beta}$, while the power allocations for slots $i=\{1, \ldots, i^*\}$ are strictly decreasing and strictly higher than $(1-\alpha) \frac{T_c-T_e}{\beta}$. Hence, we initialize $i^*=D$ and search for a solution for the powers satisfying the KKTs. If we obtain a solution then we stop and this is the optimal solution. Otherwise, we decrease $i^*$ by one and repeat the search.

\section{Implicit Temperature Constraint} \label{sec_implicit}
We now consider the case when the dynamic range of the temperature increases. In this case, we need to consider the change in the thermal noise of the system due to temperature changes.
The thermal noise is linearly proportional to the temperature \cite[Chapter 11]{gray1990analysis}. The problem can be written as:
\begin{align}
\label{prob_channel_dependent}
\max_{\mathbf{P} \geq \mathbf{0}} \quad & \sum_{i=1}^{D} \frac{1}{2}  \log \left(1+\frac{P_i}{c T_{i-1} +\sigma^2} \right)     \nonumber \\
\mbox{s.t.} \quad
& \sum_{i=1}^{k}  P_{i} \leq \sum_{i=1}^{k} E_i , \ \forall k
\end{align}
where $c$ is the proportionality constant between the thermal noise and the temperature. In this setting, the noise variance in each slot is determined by the value of the temperature at the beginning of the slot.
Using (\ref{eq_temp_in_terms_of_power}) in (\ref{prob_channel_dependent}), the problem can now be written in terms of only transmission powers as follows:
\begin{align}\label{prob_usage_dependent_powers_only}
\max_{\mathbf{P}\geq \mathbf{0}} \quad & \sum_{i=1}^{D} \frac{1}{2}  \log \left(1+\frac{P_i}{c \left(\beta \sum_{k=1}^{i-1} \alpha^{i-1-k} P_{k} +T_e\right) +\sigma^2} \right)     \nonumber \\
\mbox{s.t.} \quad &  \sum_{i=1}^{k}  P_{i} \leq \sum_{i=1}^{k} E_i , \ \forall k
\end{align}
where we define $\mbox{SINR}_i \triangleq \frac{P_i}{c\beta\sum_{k=1}^{i-1}\alpha^{i-1-k}P_k +cT_e + \sigma^2}$. In what follows, in order to simplify the notation, we assume without loss of generality that $c\beta=1$ and we define $\Gamma_j \triangleq \frac{cT_e+\sigma^2}{\alpha^{j}}$. Therefore, $\mbox{SINR}_i$ inside the $\log$ in (\ref{prob_usage_dependent_powers_only}) becomes $\mbox{SINR}_i = \frac{P_i}{\sum_{k=1}^{i-1}\alpha^{i-1-k}P_k +\Gamma_0}$.

The problem in this form highlights the effect of previous transmissions on subsequent slots. The transmission power at time $i$ appears as an \emph{interfering term} at slot indices greater than $i$ with an exponentially decaying weight due to the \emph{filtering} in the temperature. Using (\ref{eq_temp_in_terms_of_power}), the maximum temperature the system can reach is equal to $T_{max} \triangleq \beta \sum_{i=1}^{D} E_i +T_e$. This occurs when the transmitter transmits all its energy arrivals in the last slot. The value of $T_{max}$ is useful in determining the maximum possible temperature for the system. As we show, in the low SINR case in Section \ref{sec_low_SINR}, the optimal power allocation results in system temperature equal to $T_{max}$.

The problem in (\ref{prob_usage_dependent_powers_only}) is non-convex and determining the global optimal solution is generally a difficult task. Next, we adapt the signomial programming based iterative algorithm in \cite{chiang2007power} for the energy harvesting case. This algorithm provably converges to a local optimum point. The problem in (\ref{prob_usage_dependent_powers_only}) can be written in the following equivalent signomial minimization problem
\begin{align}\label{prob_sign}
\min_{\mathbf{P}\geq \mathbf{0}} \quad & \prod_{i=1}^{D}  \left( \frac{  \sum_{k=1}^{i-1} \alpha^{i-1-k} P_{k} +\Gamma_0}{ \sum_{k=1}^{i-1} \alpha^{i-1-k} P_{k} +\Gamma_0+ P_i} \right)     \nonumber \\
\mbox{s.t.} \quad &  \sum_{i=1}^{k}  P_{i} \leq \sum_{i=1}^{k} E_i , \ \forall k
\end{align}
The objective function in (\ref{prob_sign}) is a signomial function which is a ratio between two posynomials. Note also that the energy harvesting constraints in (\ref{prob_sign}) are posynomials in $P_i$.

In each iteration we approximate the objective by a posynomial. We do this by approximating the posynomial in the denominator by a monomoial. Appropriate choice of an approximation which satisfies the conditions in \cite{marks1978technical} guarantees convergence to a local optimal solution. Let us denote the posynomial in the $i$th denominator evaluated using a power vector $\mathbf{P}$ by $u_i(\mathbf{P})$, i.e., we have
\begin{align}
u_i(\mathbf{P})& \triangleq  \sum_{k=1}^{i + 1} v^i_k(\mathbf{P}) = \sum_{k=1}^{i - 1} \alpha^{i - 1 - k} P_{k}  +  P_i  + \Gamma_0
\end{align}
where for $k=\{1,\ldots,i-1\}$ we have $v^i_k(\mathbf{P})= \alpha^{i-1-k} P_{k}$, $v^i_i(\mathbf{P})=P_i $ and $v^i_{i+1}(\mathbf{P})=\Gamma_0$.

Using the arithmetic-geometric mean inequality, we approximate each posynomial by a monomial as follows:
\begin{align}
u_i(\mathbf{P}) \geq \left(\prod_{k=1}^{i - 1} \left(\frac{ \alpha^{i - 1 - k} P_{k}}{\theta^i_k}   \right) ^{\theta^i_k}\right)   \left(\frac{P_{i}}{\theta^i_i}\right)^{\theta^i_i}   \left(\frac{\Gamma_0}{\theta^i_{i + 1}} \right)^{\theta^i_{i + 1}}
\end{align}
where $\sum_{k=1}^{i+1} \theta^i_k=1$ for all $i=\{1,\ldots, D\}$.

We now solve the problem in (\ref{prob_sign}) iteratively. First, we initialize the power allocation to any feasible power allocation $\mathbf{P}^0$. Then, we approximate the posynomials $u_i(\mathbf{P}^0)$ using the arithmetic-geometric mean inequality shown above. In each iteration $j$, where the power allocation is $\mathbf{P}^j$, we choose $\theta^i_k$ as a function of the posynomials and the current power allocation as follows:
\begin{align}
\theta^i_k(\mathbf{P}^j) = \frac{v^i_k(\mathbf{P}^j) }{u_i(\mathbf{P}^j)}
\end{align}
which satisfies $\sum_{k=1}^{i+1} \theta^i_k(\mathbf{P}^j)=1$. This choice of $\theta^i_k(\mathbf{P}^j)$ guarantees that the iterations converge to a KKT point of the original problem \cite{marks1978technical}. In particular, for each iteration this is a geometric program and as required by \cite{marks1978technical}, this can be transformed into a convex problem; see also \cite{chiang2007power}.

The above iterative approach converges to a local optimal solution. Achieving the global optimal solution is of exponential complexity. Alternatively, to get to the optimal solution, an approach introduced in \cite{chiang2005geometric} can be used. This approach solves the following problem iteratively:
\begin{align}\label{prob_heurestic}
\min_{\mathbf{P}\geq \mathbf{0},t} \quad & t     \nonumber \\
\mbox{s.t.} \quad &  O(\mathbf{P})  \leq t  \nonumber\\ & t\leq \frac{t_0}{\alpha} \nonumber\\ & \sum_{i=1}^{k}  P_{i} \leq \sum_{i=1}^{k} E_i, \ \forall k
\end{align}
where $O(\mathbf{P})$ is the objective function of (\ref{prob_sign}) and $\alpha$ is chosen to be a number which is slightly more than 1 and $t_0$ can be initialized to be the solution of problem (\ref{prob_sign}) and then updated as the optimal solutions resulting from (\ref{prob_heurestic}).

This completes our treatment of the general problem for the case of implicit temperature constraints. In the following two subsections, we consider the two special cases of low and high SINR, where we are able to provide more structural solutions.

\subsection{Low SINR Case}\label{sec_low_SINR}
The low SINR case occurs when the incoming energies are small with respect to the noise variance. In this case, an approximation to the logarithm function in the objective function is the linear function, i.e., $\log(1+x) \approx x$. Hence, the objective function of (\ref{prob_usage_dependent_powers_only}) can be written as follows, c.f. \cite[equation (14)]{ulukus2000throughput}:
\begin{align}\label{obj_linear}
\sum_{i=1}^{D}  \frac{P_i}{  \sum_{k=1}^{i-1} \alpha^{i-1-k} P_{k} +\Gamma_0}
\end{align}
We next show that the optimal power allocation dictates that the energy is saved till the last slot and transmitted then, i.e.,
\begin{align}
P_i^* =0 , \quad i \leq D-1,  \ \ \text{and} \ \ P_D^*=\sum_{i=1}^{D} E_i
\end{align}
This can be proved by developing an upper bound as follows:
\begin{align}
\sum_{i=1}^{D}  \frac{P_i}{  \sum_{k=1}^{i-1} \alpha^{i-1-k} P_{k} +\Gamma_0} & \leq  \sum_{i=1}^{D}  \frac{P_i}{\Gamma_0}
\\ &\leq   \frac{\sum_{i=1}^{D}  E_i}{\Gamma_0}
\end{align}
and noting that this bound is achieved by the claimed power allocation.

A sufficient condition to have a low SINR regime is $\sum_{i=1}^{D}  E_i \ll {\Gamma_0}$. The temperature at the end of the communication session is equal to $T_{max}=\beta \sum_{i=1}^{D} E_i +T_e$. Also, the optimal power allocation does not need the non-causal knowledge of the energy arrival process, as all the harvested energy is used in the last slot.

\subsection{High SINR Case}\label{sec_high_SINR}
When the values of $c$ and $\sigma$ are small, SINR is high and we approximate the objective function by ignoring $1$ inside the logarithm, i.e., $\log(1+x)\approx \log(x)$. Hence, the problem in (\ref{prob_usage_dependent_powers_only}) can be written as:
\begin{align}\label{prob_high_SINR}
\max_{\mathbf{P}\geq \mathbf{0}} \quad & \sum_{i=1}^{D}   \frac{1}{2}\log \left( \frac{P_i}{  \sum_{k=1}^{i-1} \alpha^{i-1-k} P_{k} +\Gamma_0} \right)     \nonumber \\
\mbox{s.t.} \quad &  \sum_{i=1}^{k}  P_{i} \leq \sum_{i=1}^{k} E_i , \ \forall k
\end{align}
The problem in (\ref{prob_high_SINR}) has the Lagrangian:
\begin{align}\label{eq_Lag_high_SINR}
\mathcal{L}=  -\sum_{i=1}^{D}    \log \left( \frac{P_i}{  \sum_{k=1}^{i-1} \alpha^{i-1-k} P_{k} +\Gamma_0} \right)  + \sum_{k=1}^{D } \mu_k   \left(   \sum_{i=1}^{k}P_{i} -  \sum_{i=1}^{k}E_i \right)
\end{align}
Taking the derivative with respect to $P_i$ gives,
\begin{align}\label{eq_derivatives_init}
\frac{\partial \mathcal{L} }{\partial P_i}   =  - \frac{1}{P_i}    +    \sum_{j = i  +  1}^{D}    \frac{  \alpha^{j  -  1  -  i} } {   \sum_{k=1}^{j -  1} \alpha^{j - 1 - k} P_k  + \Gamma_0 }   +   \sum_{k=i}^{D}   \mu_k
\end{align}
and then equating to zero gives:
\begin{align}\label{eq_derivatives}
\frac{1}{P_i}  -  \sum_{j = i + 1}^{D} \frac{  \alpha^{j - 1 - i} } {   \sum_{k=1}^{j - 1} \alpha^{j - 1 - k} P_k  + \Gamma_0 } = \sum_{k=i}^{D} \mu_k
\end{align}

Although the problem in (\ref{prob_high_SINR}) is non-convex, it is a geometric program and we show next that any local optimal solution for this problem is globally optimal.
To show this, we consider the following equivalent problem:
\begin{align}\label{prob_high_SINR_equiv_convex}
\min_{\mathbf{x} \in \mathbb{R}^D} \quad & \sum_{i=1}^{D} \frac{1}{2}  \log \left( \frac{   \sum_{k=1}^{i-1} \alpha^{i-1-k} e^{x_k} +\Gamma_0}{e^{x_i}} \right)     \nonumber \\
\mbox{s.t.} \quad &  \sum_{i=1}^{k}  e^{x_i} \leq \sum_{i=1}^{k} E_i, \ \forall k
\end{align}
This equivalent problem is obtained by substituting $P_i = e^{x_i}$ and letting $x_i \in \mathbb{R}$.
The equivalent problem in (\ref{prob_high_SINR_equiv_convex}) is a convex optimization problem since the objective is a convex function in the form of a log-sum-exponent and the constraint set is a convex set \cite{boyd}. Hence, the KKTs are necessary and sufficient for global optimality. We show this as follows.

We first write the Lagrangian of problem (\ref{prob_high_SINR_equiv_convex}) as:
	\begin{align}\label{lag_equiv_convx}
	\mathcal{L}= - \sum_{i=1}^{D}    \log \left( \frac{e^{x_i}}{  \sum_{k=1}^{i-1} \alpha^{i-1-k} e^{x_k} + \Gamma_0} \right)  + \sum_{k=1}^{D } \nu_k \left(   \sum_{i=1}^{k}  e^{x_i}  -  \sum_{i=1}^{k}  E_i \right)
	\end{align}
	Taking the derivative with respect to $x_i$ gives,
	\begin{align}\label{eq_derivatives_init_cvx_equiv}
	\frac{\partial \mathcal{L} }{\partial x_i} =  -1 + \sum_{j = i+1}^{D} \frac{  \alpha^{j-1-i} e^{x_i} } {   \sum_{k=1}^{j-1} \alpha^{j-1-k} e^{x_k} + \Gamma_0 }  +e^{x_i} \sum_{k=i}^{D} \nu_k
	\end{align}
	which provides the following necessary condition:
	\begin{align}\label{eq_derivatives_conv}
	 e^{-x_i}   -  \sum_{j = i + 1}^{D} \frac{  \alpha^{j - 1 - i}   } {   \sum_{k=1}^{j - 1} \alpha^{j - 1 - k} e^{x_k}  + \Gamma_0 }  =  \sum_{k=i}^{D} \nu_k
	\end{align}
	Using the transformation $x_i=\log(P_i)$ and setting $\nu_i=\mu_i$, we observe that any solution of (\ref{eq_derivatives}) satisfies (\ref{eq_derivatives_conv}). Also, complementary slackness corresponding to (\ref{eq_Lag_high_SINR}) is satisfied if and only if it is satisfied by those for (\ref{lag_equiv_convx}). Since the equivalent problem in (\ref{prob_high_SINR_equiv_convex}) is convex, any solution satisfying the KKTs is global optimal and through the transformation $x_i=\log(P_i)$, $ \mu_i=\nu_i$ is also global optimal in the original problem in (\ref{prob_high_SINR}).

 The equivalent problem in (\ref{prob_high_SINR_equiv_convex}) can be solved using any convex optimization toolbox. We further note that the equivalent problem and the original problem both have unique solutions. More generally, for any fixed multipliers $\boldsymbol{\mu}$, the primal problem of minimizing the Lagrangian function in (\ref{eq_Lag_high_SINR}) has a unique solution. This follows because the Lagrangian function in (\ref{lag_equiv_convx}) is strictly convex as it is formed with strictly convex constraint functions and a convex objective function; for fixed Lagrange multipliers the Lagrangian function in (\ref{lag_equiv_convx}) is strictly convex.
	
We now focus on the KKT conditions of the original problem (\ref{prob_high_SINR}). Our ultimate goal in the following discussion is to show that the optimal solution of (\ref{prob_high_SINR}) has a power allocation which is monotone increasing in time index $i$, that is, $P_i \leq P_{i+1}$. We prove this by showing that the solution of the corresponding KKTs in (\ref{eq_derivatives}) with an arbitrary $\boldsymbol{\mu} \geq 0$ is monotone increasing in time index $i$, hence, this also follows for the optimal $\boldsymbol{\mu}^*$. We provide the proof for this fact in Appendix \ref{app_thm}. The proof is enabled by developing an algorithm with an update rule which satisfies the properties of \emph{standard interference functions} introduced in \cite{yates1995framework}. Hence, from \cite[Theorem 2]{yates1995framework}, the algorithm converges to a unique fixed point. We then show that the power allocation at this unique fixed point is monotone increasing in time. Then, from strict convexity of this problem, we know that KKTs in (\ref{eq_derivatives}) have a unique solution. Hence, our algorithm converges to the unique solution of the KKTs in (\ref{eq_derivatives}) and this solution has monotone increasing power allocations. When compared to its predecessors in \cite{jingP2P, kayaEmax, omurFade, ruiZhangEH, jingBC, jingMAC}, our method yields a more general class of problems in which optimal power allocation is monotone increasing under energy harvesting constraints. We also note that due to \cite[Lemma 3]{omur_temp_journ} and since the powers are monotone increasing, the temperature sequence $T_i^{*}$ resulting from the optimal power allocation $P_i^{*}$ is also monotone increasing.

In order to obtain the optimal solution, one has to determine the optimal Lagrange multipliers $\boldsymbol{\mu}^{*}$ and the power allocation $\mathbf{P}$. This can be done numerically by using standard techniques for constrained convex optimization. In particular, one can use projected gradient descent \cite{boyd} in the equivalent convex problem in (\ref{prob_high_SINR_equiv_convex}) to determine $\boldsymbol{\mu}^{*}$ and corresponding power allocation.

\section{Explicit and Implicit Temperature Constraints}\label{sec_noth_constraints}

In this section, we consider the case when both implicit and explicit temperature constraints are active. In this case, the temperature controls the channel quality and is also constrained by a critical level. This problem is in the following form:
\begin{align}
\label{prob_explicit_implicit_powers}
\max_{\mathbf{P} \geq \mathbf{0}} \quad & \sum_{i=1}^{D} \frac{1}{2}  \log \left(1+\frac{P_i}{  \sum_{k=1}^{i-1} \alpha^{i-1-k} P_{k} +\Gamma_0} \right)     \nonumber \\
\mbox{s.t.}   \quad &   \sum_{i=1}^{k} \alpha^{k-i} P_{i}  \leq \frac{T_c-T_e}{\beta}    \nonumber \\
& \sum_{i=1}^{k}  P_{i} \leq \sum_{i=1}^{k} E_i , \ \forall k
\end{align}
which is a non-convex optimization problem. We can tackle the challenge due to non-convexity here as we did in Section \ref{sec_implicit}. In particular, in the general SINR case, one can reach a local optimal solution for problem (\ref{prob_explicit_implicit_powers}) using the signomial programming approach described there. On the other hand, in the low SINR case, the objective function in (\ref{prob_explicit_implicit_powers}) is approximated by $\sum_{i=1}^{D}  \frac{P_i}{ \sum_{k=1}^{i-1} \alpha^{i-1-k} P_{k} +\Gamma_0}$ and it is a fractional program which can in general be mapped to a linear program.

The problem in (\ref{prob_explicit_implicit_powers}) possesses some of the properties of the problem with explicit temperature constraints only studied in Section \ref{sect_explicit}. In particular, if the temperature constraint is tight for two consecutive slots in the optimal solution, then the power level in the second slot must be equal to $\frac{T_c-T_e}{\beta} (1-\alpha)$. Additionally, when the temperature at the end of a slot hits $T_c$ for the first time, then the power in that slot must be strictly higher than $\frac{T_c-T_e}{\beta} (1-\alpha)$. We also note that the problem reduces to the case of implicit temperature constraint when the energy arrivals satisfy $\sum_{i=1}^{D} E_i \leq \frac{T_c -T_e}{\beta}$ as the explicit temperature constraint is never tight in this case.

In the high SINR case, we have $\log(1+x) \approx \log(x)$ and the problem (\ref{prob_explicit_implicit_powers}) is a geometric program which can be transformed to an equivalent convex problem. In general, the optimal power sequence does not have a monotonic structure in this case. When harvested energies are sufficiently large and the energy constraints are not binding, and if $\alpha \leq \frac{\beta \Gamma_0}{T_c-T_e + \beta \Gamma_0}$, then the optimal power sequence $P_i$ is monotone decreasing and when the temperature hits $T_c$, the power becomes constant and is equal to $\frac{(T_c-T_e)(1-\alpha)}{\beta}$.  Furthermore, under this condition the temperature is monotone increasing.
We provide the proof for these facts in Appendix \ref{apndx_f}.

\section{Numerical Results}

\begin{figure}[t]
	\centerline{\includegraphics[width=0.60\columnwidth]{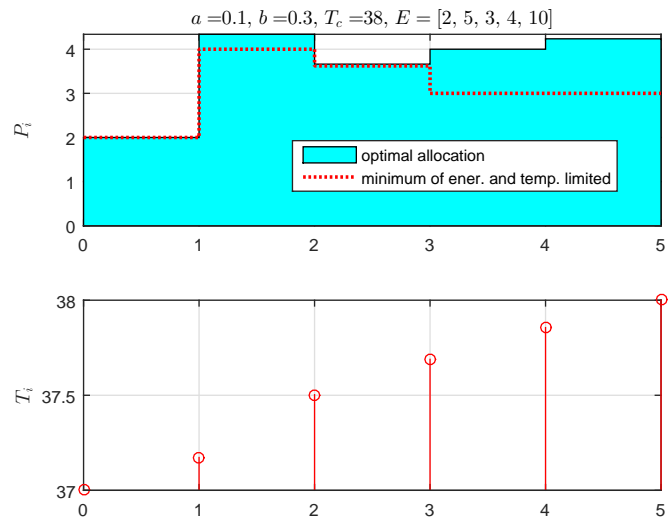}}
	\caption{Simulation for explicit temperature constraint: general case.}
	\label{res1}
\end{figure}

\begin{figure}[t]
	\centerline{\includegraphics[width=0.60\columnwidth]{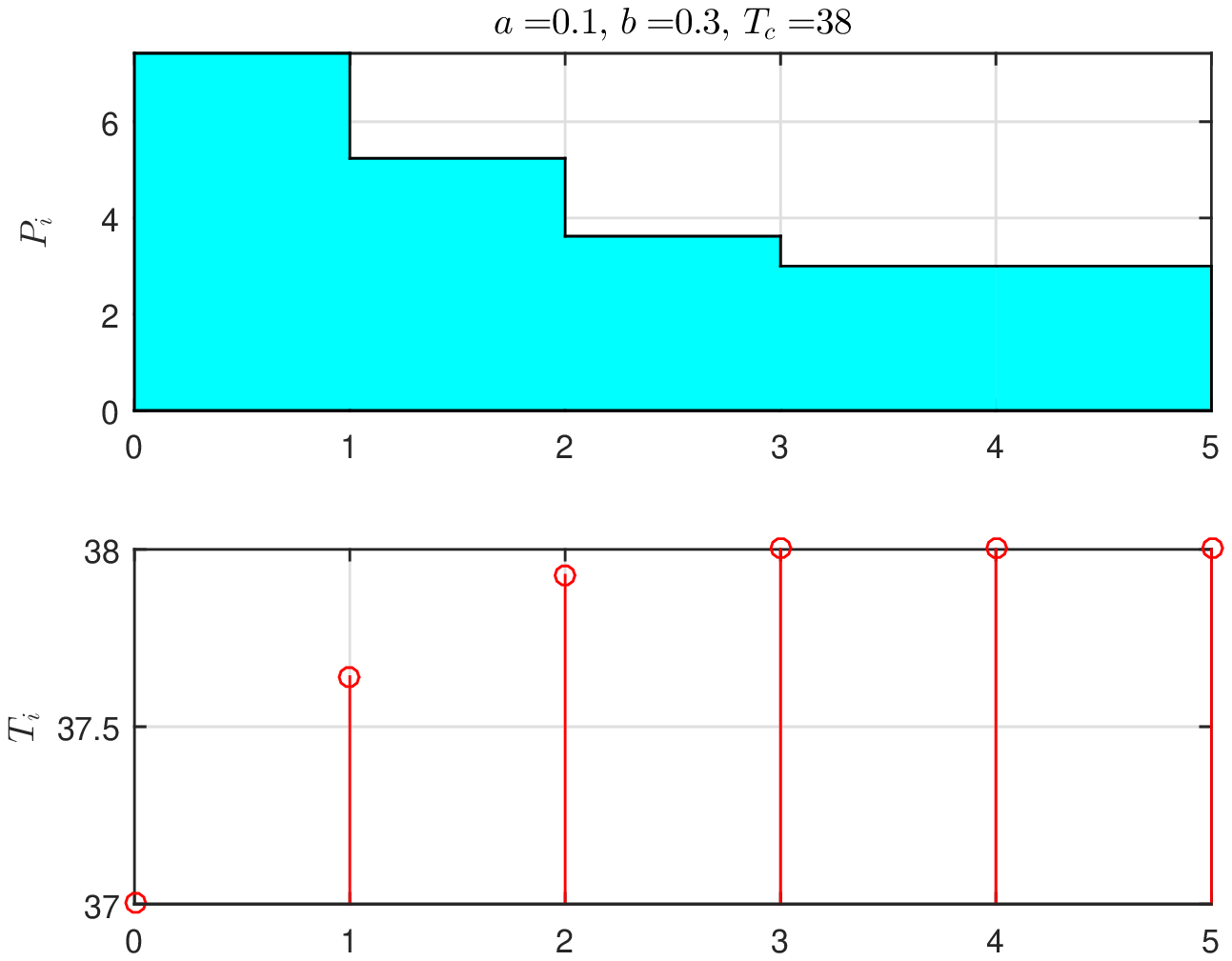}}
	\caption{Simulation for explicit temperature constraint: temperature limited case.}
	\label{res2}
\end{figure}

We first consider the explicit peak temperature constrained model considered in Section \ref{sect_explicit}. As shown in Fig. \ref{res1}, in general the power allocation does not possess any monotonicity. The optimal power allocation is close to the minimum of the power allocation of the energy and temperature limited cases.
We study the temperature limited case considered in Section \ref{sec_temp_limited} in Fig. \ref{res2}. When the temperature is strictly increasing, the power is strictly decreasing. When the temperature reaches the critical level, the power remains constant.

\begin{figure}[t]
	\centerline{\includegraphics[width=0.60\columnwidth]{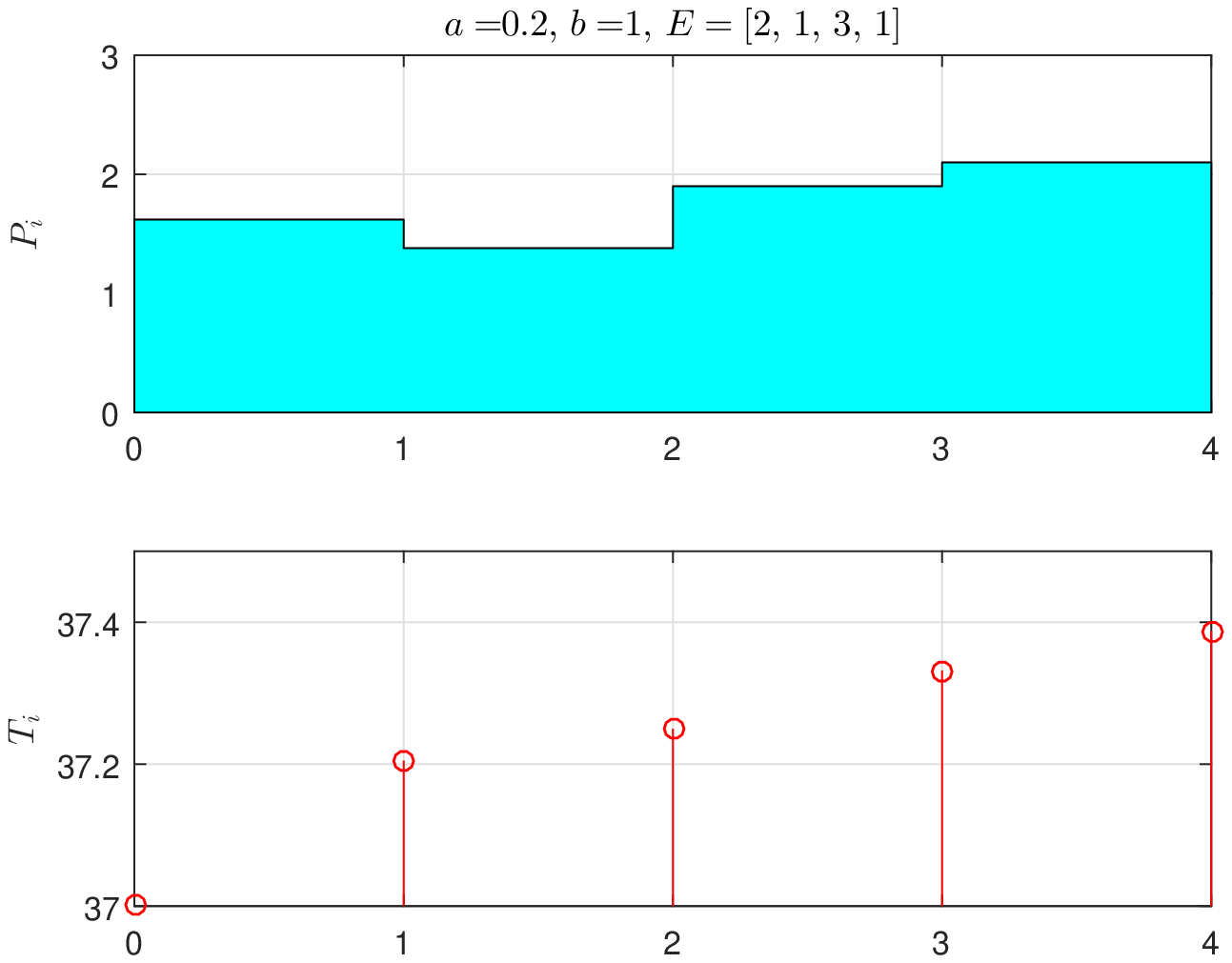}}
	\caption{Simulation for implicit temperature constraint: general case.}
	\label{res3}
\end{figure}

\begin{figure}[t]
	\centerline{\includegraphics[width=0.60\columnwidth]{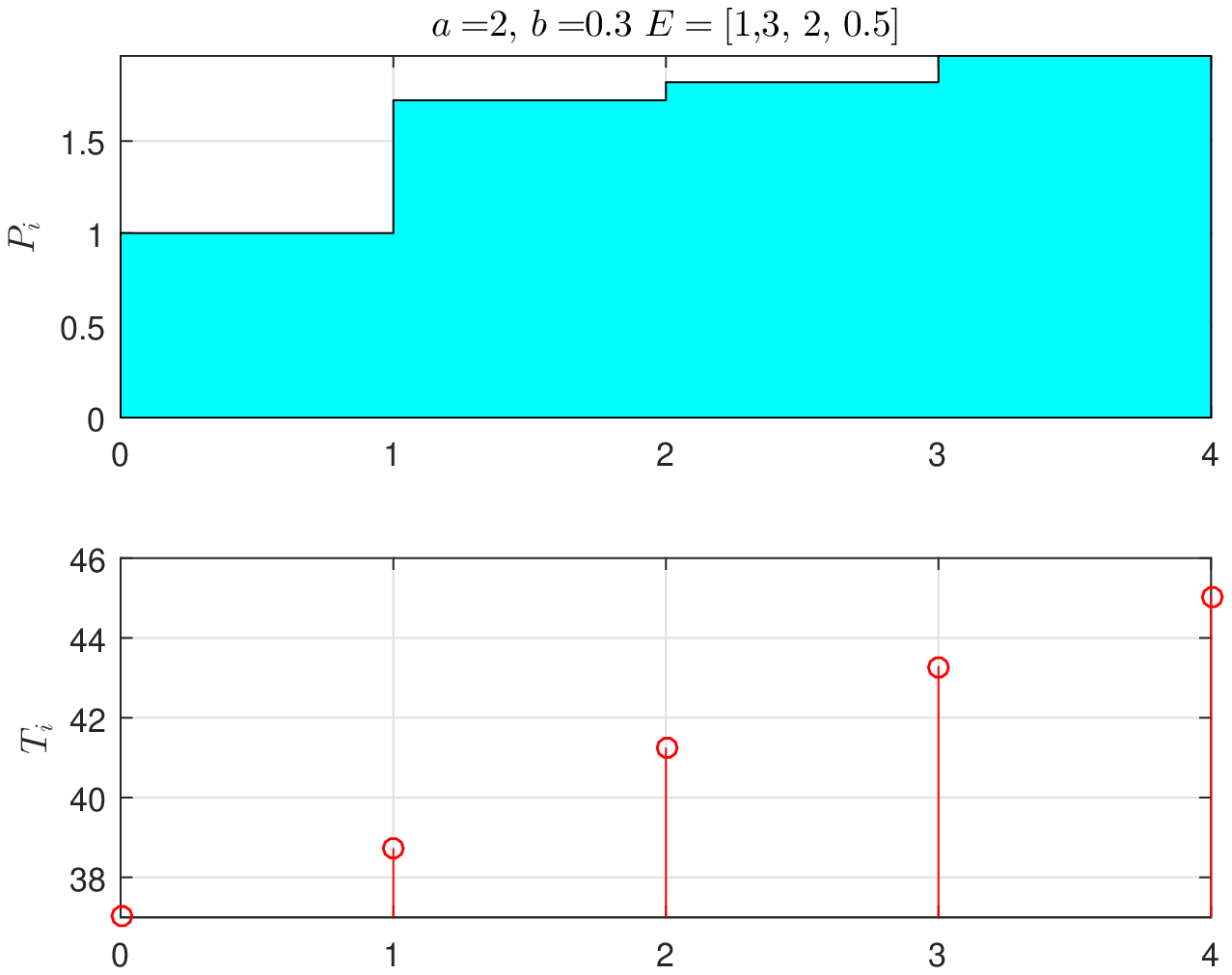}}
	\caption{Simulation for implicit temperature constraint: high SINR case.}
	\label{res5}
\end{figure}

We then consider implicit temperature constrained model considered in Section \ref{sec_implicit}. For the general SINR case, we initialize the signomial programming problem using a feasible power allocation of $P_i=\min_i E_i$ in all slots. For the case shown in Fig. \ref{res3}, the objective function takes the value $0.0895$ at the the global optimal and our experiment verifies that the single condensation method also yields $0.0895$. In general, we observe that the single condensation method gives solutions close the global optimal. We then present the high SINR case considered in Section \ref{sec_high_SINR} in Fig. \ref{res5}. We observe that the optimal power allocation is monotone increasing as proved.

\begin{figure}[t]
	\centerline{\includegraphics[width=0.60\columnwidth]{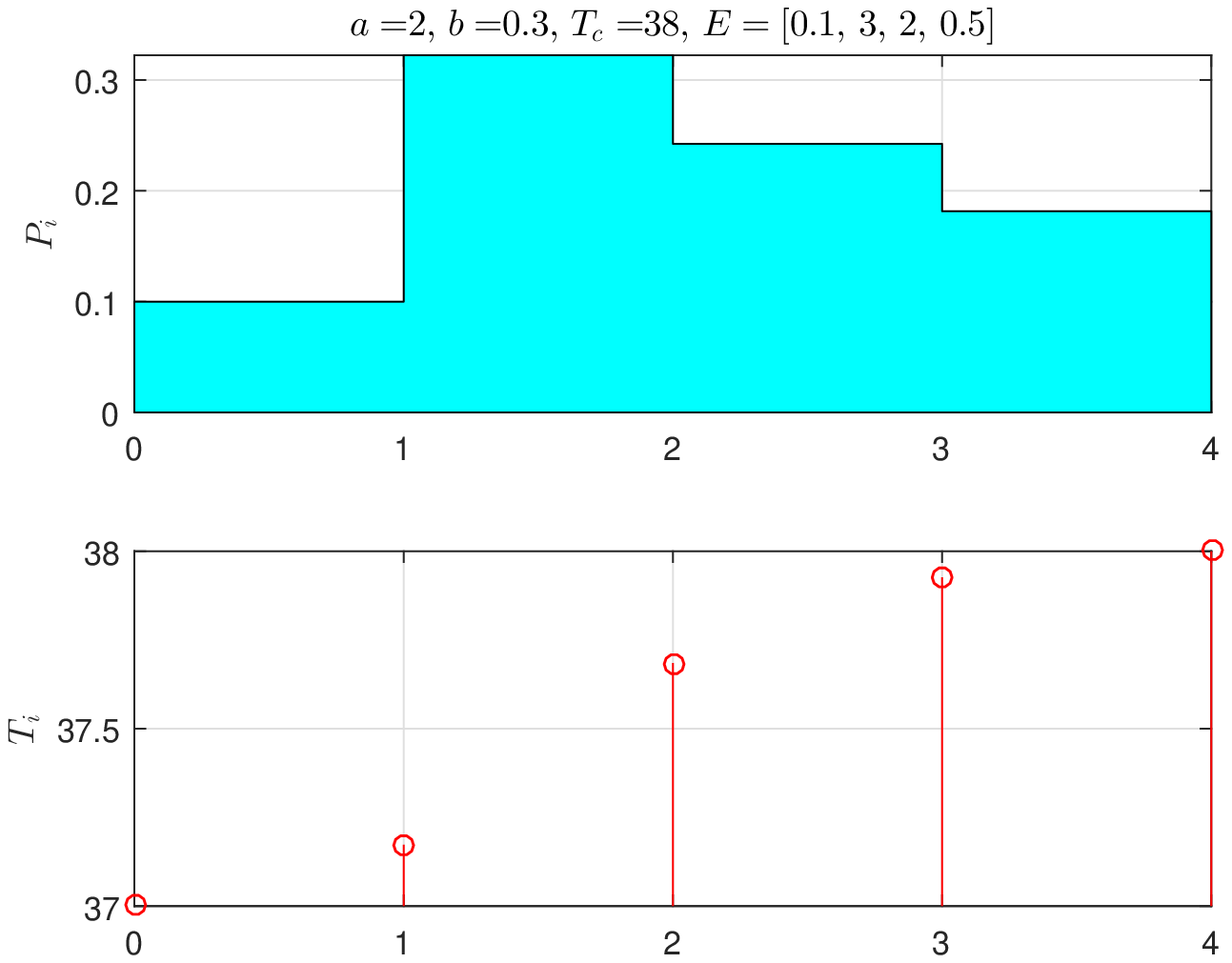}}
	\caption{Simulation for implicit and explicit temperature constraint: high SINR case.}
	\label{res6}
\end{figure}

\begin{figure}[t]
	\centerline{\includegraphics[width=0.60\columnwidth]{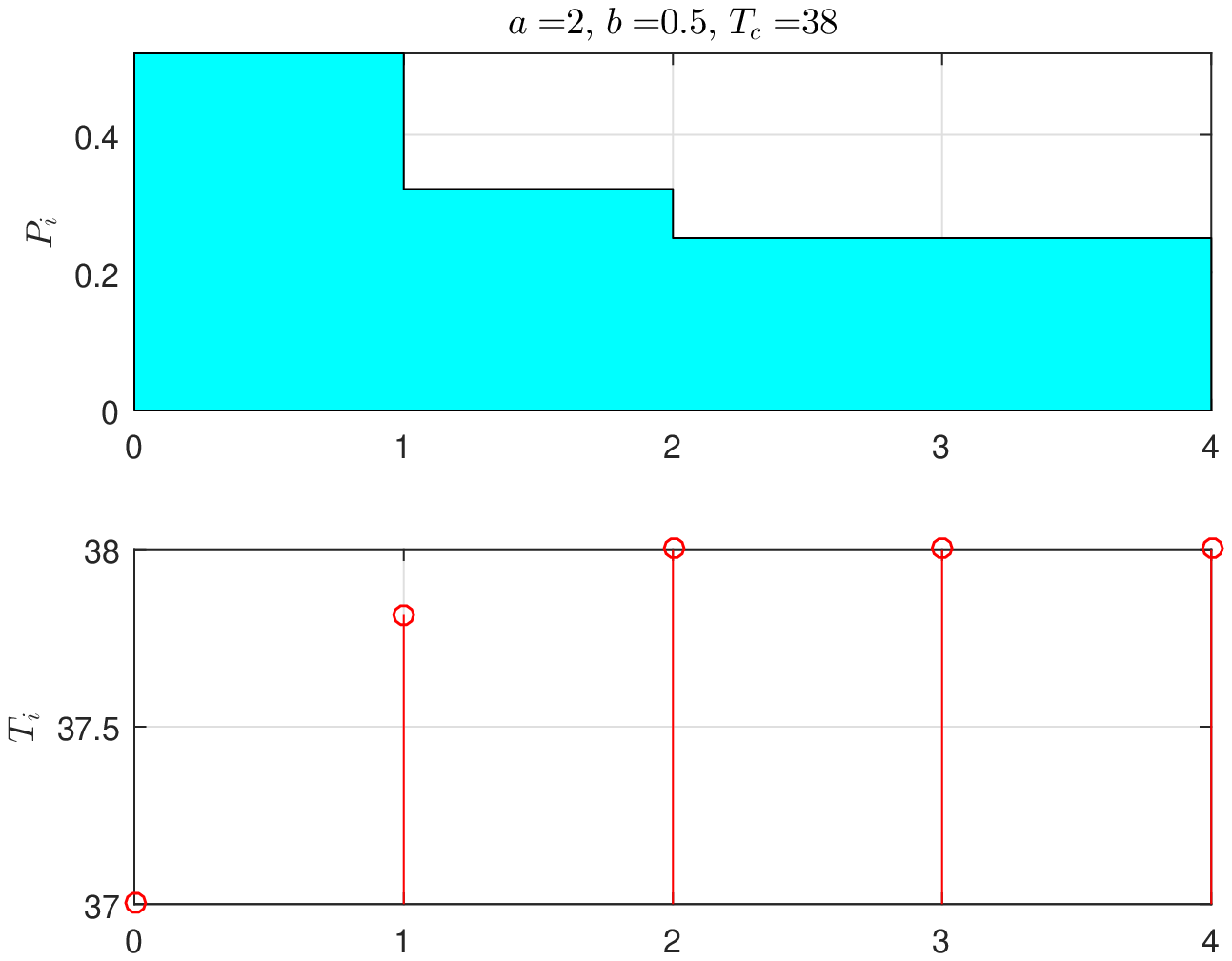}}
	\caption{Simulation for implicit and explicit temperature constraint: temperature limited high SINR case.}
	\label{res7}
\end{figure}

Next, we study the case when implicit and explicit temperature constraints are simultaneously active as considered in Section \ref{sec_noth_constraints}. For the high SINR case, unlike the implicit temperature constrained case, we observe in Fig. \ref{res6} that the power sequence does not possess a monotonic structure. We then study the high SINR case when the system is temperature limited in Fig. \ref{res7}. The optimal power allocation is monotone decreasing, corresponding temperature is monotone increasing and the power is constant when the temperature reaches the critical level.

\section{Conclusions}
We considered explicit and implicit temperature constraints in a single-user energy harvesting communication system in discrete time. Under explicit temperature constraints, the temperature is imposed to be less than a critical level. In this case, we studied optimal power allocation for multiple energy arrivals. For the temperature limited regime, we showed that the optimal power sequence is monotone decreasing while the temperature of the system is monotone increasing. Next, we considered an implicit temperature constraint where the temperature level affects channel quality. We studied the general case as well as the high and low SINR cases. In the low SINR case, we showed that the optimal allocation dictates the transmitter to save its harvested energy till the last slot and transmit all the harvested energy then. In the high SINR case, we observed that the problem is a geometric program and we expanded upon its equivalent convex version to show that optimal power allocation is monotone increasing in time. Finally, we considered the case in which implicit and explicit temperature constraints are simultaneously active. We identified a sufficient condition on the system parameters that results in a monotone decreasing optimal power allocation. Our current investigation leaves several directions to purse in future research, such as, optimal power allocation for the finite battery case; online power allocation under explicit and implicit temperature constraints; explicit and implicit temperature constraints in multi-user settings such as broadcast and multiple access channels.

\appendices

\section{Proof of the monotonicity of optimal power allocation of Problem (\ref{prob_high_SINR})}\label{app_thm}
In this appendix, we present the proof for the monotonicity of the optimal power allocation for the problem in (\ref{prob_high_SINR}).
We rewrite its KKTs in (\ref{eq_derivatives}) as follows:
\begin{align}\label{eq_derivatives_rewritten}
\frac{1}{P_i} =   \sum_{k=i}^{D} \mu_k +  \sum_{j = i + 1}^{D} \frac{  \alpha^{j - 1 - i} } {   \sum_{k=1}^{j - 1} \alpha^{j - 1 - k} P_k  + \Gamma_0 }
\end{align}
Based on this equation, for a fixed $\boldsymbol{\mu}$, we now define an update rule to solve for the power allocation $P_i$ iteratively as follows:
\begin{align}\label{eq_update_rule}
P_i(  \mathbf{P}) &\triangleq \frac{1}{\sum_{k=i}^{D} \mu_k + \sum_{j = i + 1}^{D} \frac{  \alpha^{j - 1 - i} } {   \sum_{k=1}^{j - 1} \alpha^{j - 1 - k}  P_k  + \Gamma_0 }   }
\end{align}
where the function $P_i(  \mathbf{P})$ calculates the updated power $P_i$ when the powers are equal to $\mathbf{P}$.
The algorithm proceeds as follows: We first initialize the power allocation with any arbitrary non-negative power allocation $\mathbf{P}^0$, where the superscript denotes the iteration index. We then substitute with $\mathbf{P}^0$ in (\ref{eq_update_rule}) to obtain the new power allocation $\mathbf{P}^1$, where $\mathbf{P}^1 \triangleq (P_1(  \mathbf{P}^0), \ldots, P_D(  \mathbf{P}^0))$. Similarly, we use the powers $\mathbf{P}^1$ to obtain the updated powers $\mathbf{P}^2$, and repeat this process. We show next that this algorithm converges to a unique fixed point.

To show that these updates converge to a unique fixed point, we first present the following definition of a \emph{standard interference function} \cite{yates1995framework}:
\begin{definition}
	Interference function $I(\mathbf{P})$ is standard if for all $\mathbf{P}\geq 0$ the following properties are satisfied:
	\begin{itemize}
		\item Positivity $I(\mathbf{P})>0$.
		\item Monotonicity: If $\mathbf{P}\geq \mathbf{P}'$, then $I(\mathbf{P})\geq I(\mathbf{P}')$.
		\item Scalability: For all $\theta>1$, $\theta I(\mathbf{P})\geq  I(\theta \mathbf{P})$.
	\end{itemize}
\end{definition}

Now, we want to show that the update rule $I(\mathbf{P})= \left(P_1(\mathbf{P}), P_2(\mathbf{P}), \dots, P_D(\mathbf{P}) \right)$ is a standard function, i.e., it satisfies the three properties above.

The positivity property follows from,
\begin{align}
P_i(\mathbf{P}) \geq \frac{1}{\sum_{j=i}^{D} \mu_j} \geq \frac{1}{\mu_D} >0
\end{align}
where $\mu_D>0$ follows from (\ref{eq_derivatives}) with $i=D$ and since the power $P_D$ is finite due to the finite energy constraint.

The monotonicity property follows since the denominator of $P_i(\mathbf{P})$ is a decreasing function of the powers, and hence, $P_i(\mathbf{P})$ is an increasing function of the powers.

The scalability property follows from the following for $\theta>1$,
\begin{align}
P_i(\theta  \mathbf{P}) &= \frac{1}{\sum_{k=i}^{D} \mu_k + \sum_{j = i + 1}^{D} \frac{  \alpha^{j - 1 - i} } {   \sum_{k=1}^{j - 1} \alpha^{j - 1 - k} \theta P_k  + \Gamma_0 }   } \\
& = \frac{\theta}{\theta \sum_{k=i}^{D} \mu_k + \sum_{j = i + 1}^{D} \frac{  \alpha^{j - 1 - i} } {   \sum_{k=1}^{j - 1} \alpha^{j - 1 - k}   P_k  + \frac{\Gamma_0}{\theta} }   } \\
& < \frac{\theta}{  \sum_{k=i}^{D} \mu_k + \sum_{j = i + 1}^{D} \frac{  \alpha^{j - 1 - i} } {   \sum_{k=1}^{j - 1} \alpha^{j - 1 - k}   P_k  +  \Gamma_0  }   }  \\
& =\theta P_i(\mathbf{P})
\end{align}
This completes the proof that $I(\mathbf{P})= \left(P_1(\mathbf{P}), P_2(\mathbf{P}), \dots, P_D(\mathbf{P}) \right)$ in (\ref{eq_update_rule}) is a standard interference function.

From \cite[Theorem 2]{yates1995framework}, we now conclude that the algorithm in (\ref{eq_update_rule}) converges to a \emph{unique} fixed point. From the equivalence of (\ref{prob_high_SINR}) to the strictly convex problem in (\ref{prob_high_SINR_equiv_convex}), we know that there is only one unique solution to the equations in (\ref{eq_derivatives}), and hence, the algorithm in (\ref{eq_update_rule}) converges to the unique power allocation which solves the KKTs in (\ref{eq_derivatives}).

It now remains to show that at this unique fixed point, the power allocation is monotone increasing in time. We prove this by showing that if we begin with any arbitrary monotone increasing power allocation, the update algorithm retains this ordering for the power allocation in each iteration, and hence, in the limit. To show this, let us assume that we have an arbitrary power vector $\mathbf{P}$ which satisfies $P_i \leq P_{i+1}$ for all $i=\{1,\ldots, D-1\}$. We want to show that $P_i(\mathbf{P}) \leq P_{i+1}(\mathbf{P})$. This follows from:
\begin{align}
P_{i+1}(  \mathbf{P}) &= \frac{1}{\sum_{k=i+1}^{D} \mu_k + \sum_{j = i + 2}^{D} \frac{  \alpha^{j - 2-i} } {   \sum_{k=1}^{j - 1} \alpha^{j - 1 - k}  P_k  + \Gamma_0 }   } \label{eq_nondec1}\\
& \geq \frac{1}{\sum_{k=i}^{D} \mu_k + \sum_{j = i + 2}^{D} \frac{  \alpha^{j - 2- i} } {   \sum_{k=1}^{j - 1} \alpha^{j - 1 - k}  P_k  + \Gamma_0 }   } \label{eq_nondec2}\\
& \geq \frac{1}{\sum_{k=i}^{D} \mu_k + \sum_{j = i + 2}^{D} \frac{  \alpha^{j - 2- i} } {   \sum_{k=2}^{j - 1} \alpha^{j - 1 - k}  P_k  + \Gamma_0 }   } \label{eq_nondec3}\\
& = \frac{1}{\sum_{k=i}^{D} \mu_k + \sum_{j = i + 2}^{D} \frac{  \alpha^{j - 2- i} } {   \sum_{k=1}^{j - 2} \alpha^{j - 2 - k}  P_{k+1}  + \Gamma_0 }   }   \label{eq_nondec4}\\
& \geq \frac{1}{\sum_{k=i}^{D} \mu_k + \sum_{j = i + 2}^{D} \frac{  \alpha^{j - 2- i} } {   \sum_{k=1}^{j - 2} \alpha^{j -2- k}  P_{k} + \Gamma_0 }   }
\label{eq_nondec5}\\
& \geq \frac{1}{\sum_{k=i}^{D} \mu_k + \sum_{j = i + 1}^{D-1} \frac{  \alpha^{j -1- i} } {   \sum_{k=1}^{j - 1} \alpha^{j -1- k}  P_{k} + \Gamma_0 }   } \label{eq_nondec6}\\
& \geq \frac{1}{\sum_{k=i}^{D} \mu_k + \sum_{j = i + 1}^{D} \frac{  \alpha^{j -1- i} } {   \sum_{k=1}^{j - 1} \alpha^{j -1- k}  P_{k} + \Gamma_0 }   } \label{eq_nondec7}\\
&= P_{i}(  \mathbf{P})\label{eq_nondec8}
\end{align}
where (\ref{eq_nondec2}) follows by adding the non-negative Lagrange multiplier $\mu_i$ in the denominator, (\ref{eq_nondec3}) follows by neglecting positive terms in the denominator in the second term of the denominator, (\ref{eq_nondec4}) follows by replacing $\sum_{k=2}^{j-1}$ by $\sum_{k=1}^{j-2}$ and changing the indices inside the summation accordingly, (\ref{eq_nondec5}) follows since we have $P_{k} \leq P_{k+1}$, (\ref{eq_nondec6}) follows by replacing $\sum_{j=i+2}^{D}$ by $\sum_{j=i+1}^{D-1}$ and changing the indices inside the summation accordingly, and (\ref{eq_nondec7}) follows by adding a positive term in the denominator.
Since in each iteration the power is monotone increasing, the power allocation will also be monotone increasing at the fixed point.

\section{Proof of the monotonicity of the optimal power allocation \\ of problem (\ref{prob_explicit_implicit_powers}) when $\alpha \leq \frac{\beta \Gamma_0}{T_c-T_e + \beta \Gamma_0}$ }\label{apndx_f}
	
	We start the proof by noting that the KKT conditions and the complementary slackness conditions for the problem in (\ref{prob_explicit_implicit_powers}) are necessary and sufficient for optimality.
	
	Let us now define $i_1^*$ as the first slot at which the temperature hits $T_c$. Since $\mu_k=0$ for $k=1, \ldots, i_1^{*}-1$, KKT conditions in the integer interval $[1:i_1^{*}]$ are in the following form:
	\begin{align} \label{tw}
	\frac{1}{P_i} -  \sum_{j=i+1}^{D} \frac{1}{\sum_{k=1}^{j-1}\alpha^{i-k}P_k + \Gamma_{j-1-i}} = \alpha^{-i}W, \quad i=1, \ldots, i_1^{*}
	\end{align}
	where $W \triangleq \sum_{k=i_1^{*}}^D \alpha^{k}\mu_k$. By comparing (\ref{tw}) for $i$ and $i+1\leq i_1^*$, we have:
	\begin{align}\label{tww}
	\frac{1}{P_i} -  \frac{1}{\sum_{k=1}^{i}\alpha^{i-k}P_k +\Gamma_0} = \frac{\alpha}{P_{i+1}}
	\end{align}
	We now rewrite (\ref{tww}) as follows:
	\begin{align}
	P_{i+1}&=\alpha\frac{\sum_{k=1}^{i-1}\alpha^{i-k}P_k + P_i + \Gamma_0}{\sum_{k=1}^{i-1}\alpha^{i-k}P_k + \Gamma_0}P_i \\ &= \alpha\left(1+\frac{P_i }{\sum_{k=1}^{i-1}\alpha^{i-k}P_k + \Gamma_0}\right)P_i \label{kk}
	\end{align}
	Now, due to the temperature constraints, we have $P_i \leq \frac{T_c-T_e}{\beta}$ for all $i$ and $\frac{P_i }{\sum_{k=1}^{i-1}\alpha^{i-k}P_k +\Gamma_0} \leq \frac{T_c-T_e}{\beta \Gamma_0}$. Hence, under the assumed condition on $\alpha$, we have
	\begin{align}
	\alpha\left(1+\frac{P_i }{\sum_{k=1}^{i-1}\alpha^{i-k}P_k +\Gamma_0}\right) \leq 1
	\end{align}
	This proves that $P_{i+1} \leq P_i$ for all $i \in [1:i_1^{*}-1]$, i.e., the optimal power allocation is non-increasing in the slots $\{1,\ldots,i_1^*\}$.
	
	Now, if the temperature drops below $T_c$ after slot $i_1^*$, say at slot $i_2^*$, the KKT conditions will have the form identical to (\ref{tw}) in the interval $[i_1^{*}+1:i_2^{*}]$. Following the steps, we have that $P_{i+1}\leq P_i$ for $[i_1^{*}+1:i_2^{*}]$, i.e., the optimal power allocation is non-increasing in the slots $\{i_1^*+1,\ldots,i_2^*\}$. It remains to show that the power allocation is also non-increasing between slots $i_1^*$ and $i_1^*+1$. Note that it follows that the power in slot $i_1^*$ is strictly higher than $\frac{(T_c-T_e)(1-\alpha)}{\beta}$, while in slot $i^*_1+1$ the power can be no larger than $\frac{(T_c-T_e)(1-\alpha)}{\beta}$ as otherwise this violates the temperature constraint. Hence, the power allocation between slots $i_1^*$ and $i_1^*+1$ is non-increasing also. This concludes the proof of the first part. The proof of the monotonicity of the resulting temperature follows similar to (\ref{temp_inc_eq_1})-(\ref{temp_inc_eq_5}).

\end{document}